%% file: Lambdac_to_SKP_SKPP.tex
\documentclass[a4paper,10pt,twoside]{cpc-hepnp}
\usepackage{CJK,upgreek,fancyhdr}
\usepackage{subfigure}
\usepackage{multicol}
\usepackage{booktabs}
\usepackage{amsmath}
\usepackage{amsfonts,amssymb,bm,mathrsfs,bbm,amscd}
\usepackage{lastpage}
\usepackage[pdftex]{graphicx}
\usepackage{multirow}
\DeclareGraphicsExtensions{.pdf,.jpeg,.png}
\usepackage{epstopdf}
\usepackage[english]{babel}
\usepackage[colorlinks=false,pdfpagemode=FullScreen,,setpagesize=off,pdfborder={0 0 0}]{hyperref}
\usepackage{overpic}
\usepackage{lineno}
\usepackage{color}
\usepackage{diagbox}
\usepackage{adjustbox}

\usepackage{tikz-feynman}
\tikzfeynmanset{compat=1.1.0}

\newcommand{\ifb}{\mathrm{fb}^{-1}}

\newcommand{\ee}{e^+e^-}
\newcommand{\pip}{\pi^+}
\newcommand{\pim}{\pi^-}
\newcommand{\mev}{\,\mathrm{MeV}}

\begin{document}
\begin{CJK*}{UTF8}{gkai}
% \linenumbers

\fancyhead[c]{\small Chinese Physics C~~~Vol. xx, No. x (2024) xxxxxx} 
\fancyfoot[C]{\small 010201-\thepage}
\footnotetext[0]{Received xxxx October xxxx}

\title{Search for the Cabibbo-suppressed decays $\Lambda_c^{+}\to\Sigma^0K^{+}\pi^{0}$ and $\Lambda_c^{+}\to\Sigma^0K^{+}\pi^{+}\pi^{-}$
\thanks{
The BESIII Collaboration thanks the staff of BEPCII and the IHEP computing center for their strong support. This work is supported in part by National Key R\&D Program of China under Contracts Nos. 2020YFA0406300, 2020YFA0406400, 2023YFA1606000; National Natural Science Foundation of China (NSFC) under Contracts Nos. 12205141,  11635010, 11735014, 11935015, 11935016, 11935018, 12025502, 12035009, 12035013, 12061131003, 12192260, 12192261, 12192262, 12192263, 12192264, 12192265, 12221005, 12225509, 12235017, 12361141819; Natural Science Foundation of Hunan Province (Contracts No.2024JJ2044); the Chinese Academy of Sciences (CAS) Large-Scale Scientific Facility Program; the CAS Center for Excellence in Particle Physics (CCEPP); Joint Large-Scale Scientific Facility Funds of the NSFC and CAS under Contract No. U1832207; 100 Talents Program of CAS; The Institute of Nuclear and Particle Physics (INPAC) and Shanghai Key Laboratory for Particle Physics and Cosmology; German Research Foundation DFG under Contract No. FOR5327; Istituto Nazionale di Fisica Nucleare, Italy; Knut and Alice Wallenberg Foundation under Contracts Nos. 2021.0174, 2021.0299; Ministry of Development of Turkey under Contract No. DPT2006K-120470; National Research Foundation of Korea under Contract No. NRF-2022R1A2C1092335; National Science and Technology fund of Mongolia; National Science Research and Innovation Fund (NSRF) via the Program Management Unit for Human Resources \& Institutional Development, Research and Innovation of Thailand under Contracts Nos. B16F640076, B50G670107; Polish National Science Centre under Contract No. 2019/35/O/ST2/02907; Swedish Research Council under Contract No. 2019.04595; The Swedish Foundation for International Cooperation in Research and Higher Education under Contract No. CH2018-7756; U. S. Department of Energy under Contract No. DE-FG02-05ER41374.}}
\maketitle
\begin{center}
\input{authorlist_2024-10-09}
\end{center}
\begin{abstract}

Utilizing 4.5 $\ifb$ of $\ee$ annihilation data collected at
center-of-mass energies ranging from 4599.53 MeV to 4698.82 MeV  by
the BESIII detector at the BEPCII collider, we search for the singly
Cabibbo-suppressed hadronic decays $\Lambda_{c}^{+}\to\Sigma^{0}
K^{+}\pi^{0}$ and $\Lambda_{c}^{+}\to\Sigma^{0}K^{+}\pip\pim$ with a
single-tag method. No significant signals are observed for both decays. The upper limits on the branching fractions at the $90\%$ confidence level are determined to be $5.0\times 10^{-4}$ for  $\Lambda_{c}^{+}\to\Sigma^{0} K^{+}\pi^{0}$  and $6.5\times 10^{-4}$ for $\Lambda_c^{+}\to\Sigma^0K^{+}\pi^{+}\pi^{-}$.
\end{abstract}


\begin{multicols}{2}

\section{Introduction}

The experimental investigation of the decays of charmed baryons plays a critical role in understanding the complex dynamics of strong and weak interactions involving heavy quarks. The charmed baryon, $\Lambda_{c}^{+}$, was first observed by the Mark II experiment in 1979~\cite{1980ee}. 
However, despite the decades of research, the sum of the known $\Lambda^+_c$ decay branching fractions (BFs) is still limited to about 70\%, with the remaining decays yet to be measured~\cite{ParticleDataGroup:2024cfk, 2021Lyu}. The hadronic decay amplitudes of $\Lambda^+_c$ include both factorizable and
nonfactorizable contributions. Unlike charmed meson decays, where
factorizable contributions are dominant due to higher emitted
energy~\cite{factorizable}, the weak hadronic decays of
$\Lambda_{c}^{+}$ are not suppressed by color or
helicity~\cite{color}. As a result, they receive significant
non-factorizable contributions, such as those from the $W$-exchange diagrams. Although extensive studies with various phenomenological models have been performed on hadronic $\Lambda^+_c$ decays, the interference between the $W$-emission and $W$-exchange amplitudes remains unclear currently. 
Therefore, it is essential to further improve the measurements of the BFs of Cabibbo-favored~(CF) and singly Cabibbo-suppressed~(SCS) decays of $\Lambda^+_c$.

Study of the three-body hadronic decays $\Lambda_{c}^{+} \to B_{n}PP'$ is an important area of research , where $B_{n}$ and $P$($P'$) denote octet baryon and pseudoscalar meson, respectively. To date, the SCS hadronic decay 
\begin{center}
\begin{minipage}[b]{.9\linewidth}
  \begin{adjustbox}{width=\linewidth}
\begin{tikzpicture}

  \begin{feynman}
    \vertex (a1) {\( c\)};
    \vertex[right=2cm of a1] (a2);
    \vertex[right=2cm of a2] (a3){\(d\)};

    \vertex[below=2em of a1] (b1) {\(d\)};
    \vertex[right=2cm of b1] (b2);
    \vertex[right=2cm of b2] (b3) {\(u\)};

    \vertex[below=1em of b3] (c1) {\( s\)};
    \vertex[below=2em of c1] (c3) {\(\overline s\)};
    \vertex at ($(c1)!0.5!(c3) - (1cm, 0)$) (c2);
    
    \vertex[below=1em of c3] (d1) {\( u\)};
    \vertex[below=2em of d1] (d3) {\(\overline u\)};
    \vertex at ($(d1)!0.5!(d3) - (1cm, 0)$) (d2);
    
    \vertex[below=7em of b1] (e1) {\(u\)};
    \vertex[right=2cm of e1] (e2);
    \vertex[right=2cm of e2] (e3) {\(u\)};
    
    \diagram* {
      {[edges=fermion]
        (a1) -- (a2) -- (a3),
        (b1) -- (b2) -- (b3),
        (e1) -- (e2) -- (e3),
      },
      (a2) -- [boson, edge label=\(W^+\)] (b2),

      (c3) -- [fermion, out=180, in=-45] (c2) -- [fermion, out=45, in=180] (c1),
      (d3) -- [fermion, out=180, in=-45] (d2) -- [fermion, out=45, in=180] (d1),
     
    };

    \draw [decoration={brace}, decorate] (e1.south west) -- (a1.north west)
          node [pos=0.5, left] {\(\Lambda_{c}^{+}\)};
    \draw [decoration={brace}, decorate] (a3.north east) -- (c1.south east)
          node [pos=0.5, right] {\(\Sigma^{0}\)};
    \draw [decoration={brace}, decorate] (c3.north east) -- (d1.south east)
         node [pos=0.5, right] {\(K^{+}\)};
    \draw [decoration={brace}, decorate] (d3.north east) -- (e3.south east)
         node [pos=0.5, right] {\(\pi^{0}\)};
  \end{feynman}
  \end{tikzpicture}
    \end{adjustbox}
  \put(-85,-15){(a)}
\end{minipage}
\begin{minipage}[b]{.9\linewidth}
  \begin{adjustbox}{width=\linewidth}
\begin{tikzpicture}

  \begin{feynman}
    \vertex (a1) {\( u\)};
    \vertex[right=2cm of a1] (a2);
    \vertex[right=2cm of a2] (a3){\(u\)};

    \vertex[below=2em of a1] (b1) {\(c\)};
    \vertex[right=2cm of b1] (b2);
    \vertex[right=2cm of b2] (b3) {\(d\)};

    \vertex[below=1em of b3] (c1) {\( s\)};
    \vertex[below=2em of c1] (c3) {\(\overline s\)};
    \vertex at ($(c1)!0.5!(c3) - (1cm, 0)$) (c2);
    
    \vertex[below=1em of c3] (d1) {\( u\)};
    \vertex[below=2em of d1] (d3) {\(\overline u\)};
    \vertex at ($(d1)!0.5!(d3) - (1cm, 0)$) (d2);
    
    \vertex[below=7em of b1] (e1) {\(d\)};
    \vertex[right=2cm of e1] (e2);
    \vertex[right=2cm of e2] (e3) {\(u\)};
    
    \diagram* {
      {[edges=fermion]
        (a1) -- (a2) -- (a3),
        (b1) -- (b2) -- (b3),
        (e1) -- (e2) -- (e3),
      },
      (b2) -- [boson, edge label=\(W^+\)] (e2),

      (c3) -- [fermion, out=180, in=-45] (c2) -- [fermion, out=45, in=180] (c1),
      (d3) -- [fermion, out=180, in=-45] (d2) -- [fermion, out=45, in=180] (d1),
      
    };

    \draw [decoration={brace}, decorate] (e1.south west) -- (a1.north west)
          node [pos=0.5, left] {\(\Lambda_{c}^{+}\)};
    \draw [decoration={brace}, decorate] (a3.north east) -- (c1.south east)
          node [pos=0.5, right] {\(\Sigma^{0}\)};
    \draw [decoration={brace}, decorate] (c3.north east) -- (d1.south east)
         node [pos=0.5, right] {\(K^{+}\)};
    \draw [decoration={brace}, decorate] (d3.north east) -- (e3.south east)
         node [pos=0.5, right] {\(\pi^{0}\)};
  \end{feynman}
  \end{tikzpicture}
      \end{adjustbox}
         \put(-85,-15){(b)}
\end{minipage}
\end{center}
\begin{center}
\begin{minipage}[b]{.9\linewidth}
  \begin{adjustbox}{width=\linewidth}
\begin{tikzpicture}

  \begin{feynman}

    \vertex (a1) {\( u\)};
    \vertex[right=2cm of a1] (a2);
    \vertex[right=2cm of a2] (a3){\(u\)};

    \vertex[below=4em of a1] (b1) {\(c\)};
    \vertex[right=2cm of b1] (b2);
    \vertex[right=2cm of b2] (b3) {\(d\)};

    \vertex[below=1em of a3] (c3) {\(\overline s\)};
    \vertex[below=2em of c3] (c1) {\( s\)};
    \vertex at ($(c1)!0.5!(c3) - (1cm, 0)$) (c2);
    
    \vertex[below=1em of b3] (d1) {\( u\)};
    \vertex[below=2em of d1] (d3) {\(\overline d\)};
    \vertex at ($(d1)!0.5!(d3) - (1cm, 0)$) (d2);
    
    \vertex[below=5em of b1] (e1) {\(d\)};
    \vertex[right=2cm of e1] (e2);
    \vertex[right=2cm of e2] (e3) {\(d\)};
    
    \diagram* {
      {[edges=fermion]
        (a1) -- (a2) -- (a3),
        (b1) -- (b2) -- (b3),
        (e1) -- (e2) -- (e3),
      },
      (b2) -- [boson,bend right, edge label=\(W^+\)] (d2),

      (c3) -- [fermion, out=-180, in=45] (c2) -- [fermion, out=-45, in=180] (c1),
      (d3) -- [fermion, out=180, in=-45] (d2) -- [fermion, out=45, in=180] (d1),
     
    };

    \draw [decoration={brace}, decorate] (e1.south west) -- (a1.north west)
          node [pos=0.5, left] {\(\Lambda_{c}^{+}\)};
    \draw [decoration={brace}, decorate] (c1.north east) -- (d1.south east)
          node [pos=0.5, right] {\(\Sigma^{0}\)};
    \draw [decoration={brace}, decorate] (a3.north east) -- (c3.south east)
         node [pos=0.5, right] {\(K^{+}\)};
    \draw [decoration={brace}, decorate] (d3.north east) -- (e3.south east)
         node [pos=0.5, right] {\(\pi^{0}\)};
  \end{feynman}
  \end{tikzpicture}
\end{adjustbox}
\put(-85,-15){(c)}
\end{minipage}
\begin{minipage}[b]{.9\linewidth}
  \begin{adjustbox}{width=\linewidth}
\begin{tikzpicture}
%\hspace*{-5cm}
  \begin{feynman}

    \vertex (a1) {\( c\)};
    \vertex[right=2cm of a1] (a2);
    \vertex[right=2cm of a2] (a3){\(s\)};
    
    \vertex[below=4em of a1] (b1) {\(d\)};
    \vertex[right=2cm of b1] (b2);
    \vertex[right=2cm of b2] (b3) {\(d\)};

    \vertex[below=1em of b3] (c3) {\( u\)};
    \vertex[below=2em of c3] (c1) {\( \overline u\)};
    \vertex at ($(c1)!0.5!(c3) - (1cm, 0)$) (c2);
    
    \vertex[above=4em of a3] (d1) {\( u\)};
    \vertex[below=2em of d1] (d3) {\(\overline s\)};
    \vertex at ($(d1)!0.5!(d3) - (1cm, 0)$) (d2);
    
    \vertex[below=5em of b1] (e1) {\(u\)};
    \vertex[right=2cm of e1] (e2);
    \vertex[right=2cm of e2] (e3) {\(u\)};
    
    \diagram* {
      {[edges=fermion]
        (a1) -- (a2) -- (a3),
        (b1) -- (b2) -- (b3),
        (e1) -- (e2) -- (e3),
      },
      (a2) -- [boson,bend left, edge label=\(W^+\)] (d2),

      (c1) -- [fermion, out=180, in=-45] (c2) -- [fermion, out=45, in=180] (c3),
      (d3) -- [fermion, out=180, in=-45] (d2) -- [fermion, out=45, in=180] (d1),
    
    };

    \draw [decoration={brace}, decorate] (e1.south west) -- (a1.north west)
          node [pos=0.5, left] {\(\Lambda_{c}^{+}\)};
    \draw [decoration={brace}, decorate] (a3.north east) -- (c3.south east)
          node [pos=0.5, right] {\(\Sigma^{0}\)};
    \draw [decoration={brace}, decorate] (d1.north east) -- (d3.south east)
         node [pos=0.5, right] {\(K^{+}\)};
    \draw [decoration={brace}, decorate] (c1.north east) -- (e3.south east)
         node [pos=0.5, right] {\(\pi^{0}\)};
  \end{feynman}
  \end{tikzpicture}
\end{adjustbox}
\put(-85,-15){(d)}
\end{minipage}
	\figcaption{ \label{fig:feynmand} The Feynman diagrams of $\Lambda_{c}^{+}\to\Sigma^{0} K^{+}\pi^{0}$: (a)~and (b)~$W$-exchange diagrams, (c)~Internal $W$-emission diagram, and (d)~External $W$-emission diagram. }
\end{center}
$\Lambda_{c}^{+} \to \Sigma^{0} K^{+}\pi^{0}$ has not been observed. This decay can proceed via the Feynman diagrams shown in Fig.~\ref{fig:feynmand}.
Predictions of the decay BF range from $0.8 \times 10^{-3}$ to $1.2
\times 10^{-3}$ with the SU(3) flavor symmetry
framework~\cite{Geng:2018upx, Cen:2019ims, 2024LMD}, under the
assumption that the $PP'$ system is in an $S$-wave state. The most
recent prediction, as reported in Ref.~\cite{2024LMD}, considers the complete effective Hamiltonian contribution. 
On the other hand, Ref.~\cite{2018MGronau} predicts the BF of
$\Lambda^+_c\to \Sigma^0K^+\pi^0$ to be $(2.1\pm0.6) \times 10^{-3}$
with the statistical isospin model. Experimentally, the BESIII
experiment searched for this decay for the first time using a double-tag method and set an upper limit on its BF at the 90\% confidence level~(C.L.) to be $1.8\times 10^{-3}$~\cite{BESIII:2023SKP}.    

Currently, there are no theoretical predictions of the four-body hadronic decay $\Lambda_c^{+} \to \Sigma^0K^{+}\pi^{+}\pi^{-}$. The BaBar experiment performed the first search for this decay and reported an upper limit on the BF ratio $\frac{\mathcal{B} (\Lambda_c^{+} \to \Sigma^0K^{+}\pi^{+}\pi^{-})}{\mathcal{B} (\Lambda_{c}^{+} \to \Sigma^{0}\pi^+)} < 2.0 \times 10^{-2}$ at the 90\% C.L.~\cite{BaBar:2006eah}.

In this work, we search for the hadronic decays $\Lambda_{c}^{+} \to \Sigma^{0} K^{+} \pi^{0}$ and $\Lambda_{c}^{+} \to \Sigma^{0} K^{+} \pi^{+} \pi^{-}$, with subsequent decay $\Sigma^{0} \to \gamma \Lambda$, utilizing 4.5~fb$^{-1}$ of $e^{+}e^{-}$ annihilation data collected at center-of-mass (c.m.) energies ranging from 4599.53 MeV to 4698.82 MeV~\cite{BESIII:2022Lambdac1, BESIII:2016Lambdac, BESIII:2022Lambdac2}. Throughout this paper, the charge-conjugate state is always implied.

\section{BESIII detector and Monte Carlo simulation}
The BESIII detector~\cite{Ablikim:2009aa} records symmetric $e^+e^-$ collisions 
provided by the BEPCII storage ring~\cite{Yu:IPAC2016-TUYA01}
in the c.m. energy range from 1.84 to  4.95~GeV,
with a peak luminosity of $1.1\times 10^{33}\;\text{cm}^{-2}\text{s}^{-1}$ 
achieved at $\sqrt{s} = 3.773\;\text{GeV}$. 
BESIII has collected large data samples in this energy region~\cite{Ablikim:2019hff}. The cylindrical core of the BESIII detector covers 93\% of the full solid angle and consists of a helium-based
 multilayer drift chamber~(MDC), a plastic scintillator time-of-flight
system~(TOF), and a CsI(Tl) electromagnetic calorimeter~(EMC),
which are all enclosed in a superconducting solenoidal magnet
providing a 1.0~T magnetic field.
The solenoid is supported by an
octagonal flux-return yoke with resistive plate counter muon
identification modules interleaved with steel. The charged-particle momentum resolution at $1~{\rm GeV}/c$ is
$0.5\%$, and the 
${\rm d}E/{\rm d}x$
resolution is $6\%$ for electrons
from Bhabha scattering. The EMC measures photon energies with a
resolution of $2.5\%$ ($5\%$) at $1$~GeV in the barrel (end-cap)
region. The time resolution in the TOF barrel region is 68~ps, while
that in the end-cap region was 110~ps. The end-cap TOF system was
upgraded in 2015 using multigap resistive plate chamber technology,
providing a time resolution of 60 ps~\cite{etof1, etof2, etof3}. About 87\% of the data used in this analysis benefits from this upgrade. 

Simulated samples, generated using the {\sc geant4}-based~\cite{geant4} Monte Carlo (MC) package, which
includes the geometric description of the BESIII detector and the detector response, are used to determine the detection efficiency and to estimate the backgrounds. The simulation includes the beam energy spread and initial state radiation (ISR) in the $e^+e^-$ annihilation, modeled with the generator {\sc
kkmc}~\cite{ref:kkmc}. For the ISR simulation, the Born cross section
line shape of $e^+e^-\to\Lambda_c^{+}\bar{\Lambda}_c^{-}$ measured by
BESIII is used~\cite{ref:BESIIILcpair}.
Signal MC samples are generated as
$\Lambda_c^{+}\to\Sigma^0K^{+}\pi^{0}$,
$\Lambda_c^{+}\to\Sigma^0\pi^{+}\pi^{0}$,
$\Lambda_c^{+}\to\Sigma^0K^{+}\pi^{+}\pi^{-}$, and
$\Lambda_c^{+}\to\Sigma^0\pi^{+}\pi^{+}\pi^{-}$, with the
$\bar{\Lambda}_c^{-}$ baryon decays inclusively. The signal decays are
produced using the phase space (PHSP) model. To calculate the
detection efficiencies, one million signal MC events are generated for each energy point, where $\Lambda_c^+$ ($\bar{\Lambda}_c^-$) decays into the signal mode, and $\bar{\Lambda}_c^-$ ($\Lambda_c^+$) decays into all possible states. Additionally, to study the peaking background, exclusive MC samples of $\Lambda_c^{+}\to\Xi^{0} K^{+}$ and $\Lambda_c^{+}\to\Lambda K^{*+}$ are generated. Inclusive MC samples consist of open-charm states, ISR production of the $J/\psi$ and $\psi(3686)$ states, and continuum processes $e^+e^-\to q\bar{q}~(q = u, d, s)$ are used to study backgrounds.
The known decay modes of charmed hadrons and charmonium states are modeled with {\sc
evtgen}~\cite{ref:evtgen,ref:RG} using BFs taken from the
Particle Data Group (PDG)~\cite{ParticleDataGroup:2024cfk}, and the remaining unknown decays
are modeled with {\sc lundcharm}~\cite{ref:lundcharm, ref:lundcharm2}. Final state radiation
from charged final-state particles is incorporated with the {\sc
photos} package~\cite{photos}.

\section{Event selection and data analysis}

Due to limited data statistics, we adopt a single-tag approach to improve signal efficiencies, where only one $\Lambda_c^{+}$ is reconstructed in each event, with no requirement on the recoil side.
To avoid potential bias and validate the analysis procedure, a blind analysis is adopted to analyze pseudodata, such as the inclusive MC sample with equivalent size as data. The real data is unblinded after the analysis procedure has been fixed.

All charged tracks are required to have a polar angle ($\theta$) range within $|\rm{cos\theta}|<0.93$, where $\theta$ is defined with respect to the $z$ axis, which is the symmetry axis of the MDC. For the charged tracks not originating from $\Lambda$ decays, the distance of closest approach to the interaction point (IP) must be less than 10\,cm along the $z$-axis, $|V_{z}|$,  and less than 1\,cm in the transverse plane, $V_{xy}$. 
Particle identification~(PID) for charged tracks combines measurements of the energy deposited in the MDC~(d$E$/d$x$) and the flight time in the TOF to form likelihoods $\mathcal{L}(h)~(h=p,K,\pi)$ for each hadron $h$ hypothesis.
Tracks are identified as protons when the proton hypothesis has the greatest likelihood ($\mathcal{L}(p)>\mathcal{L}(K)$ and $\mathcal{L}(p)>\mathcal{L}(\pi)$), while charged kaons and pions are identified by comparing the likelihoods for the kaon and pion hypotheses, $\mathcal{L}(K)>\mathcal{L}(\pi)$ and $\mathcal{L}(\pi)>\mathcal{L}(K)$, respectively.

The $\Lambda$ particles are reconstructed from a pair of oppositely charged proton and pion candidates satisfying $|V_{z}|<$ 20~cm. The same PID
requirements as mentioned before are imposed to select the proton
candidates. Other charged tracks are assigned to be $\pi$ candidate without any PID requirements. These charged tracks are constrained to originate from the common decay vertex by requiring the $\chi^{2}$ of the vertex fit to be less than 100, and the decay length is required to be greater than
twice the vertex resolution away from the IP. To ensure reconstruction reliability, the $\Lambda$ candidates are required to have an invariant mass within $1.111 < M(p\pi^{-}) < 1.121$ GeV/$c^{2}$, which corresponds to three times of the mass resolution around the known $\Lambda$ mass~\cite{ParticleDataGroup:2024cfk}.

Photon candidates are identified using isolated showers in the EMC.  The deposited energy of each shower must be more than 25~MeV in the barrel region ($|\cos \theta|< 0.80$) and more than 50~MeV in the end cap region ($0.86 <|\cos \theta|< 0.92$). To exclude showers that originate from charged tracks, the angle subtended by the EMC shower and the position of the closest charged track at the EMC must be greater than $10^\circ$ as measured from the IP. 
To suppress electronic noise and showers unrelated to the event, the difference between the EMC time and the event start time is required to be within [0, 700]\,ns.
The $\pi^{0}$ candidates are reconstructed from photon pairs with an invariant mass within $0.115 < M(\gamma\gamma) < 0.150$ GeV/$c^{2}$. To improve momentum resolution, a one-constraint kinematic fit is utilized to constrain the $M(\gamma\gamma)$ to the known $\pi^0$ mass~\cite{ParticleDataGroup:2024cfk}. Only combinations that satisfy $\chi^{2}<200$ are retained, and the refined momenta are then employed for subsequent analysis. Then, the $\Sigma^{0}$ candidates are reconstructed from the $\Lambda\gamma$ final states, with an invariant mass in the range $1.179 < M(\Lambda\gamma) < 1.203$ GeV/$c^{2}$.

\begin{figure*}
    \includegraphics[width=0.5\textwidth]{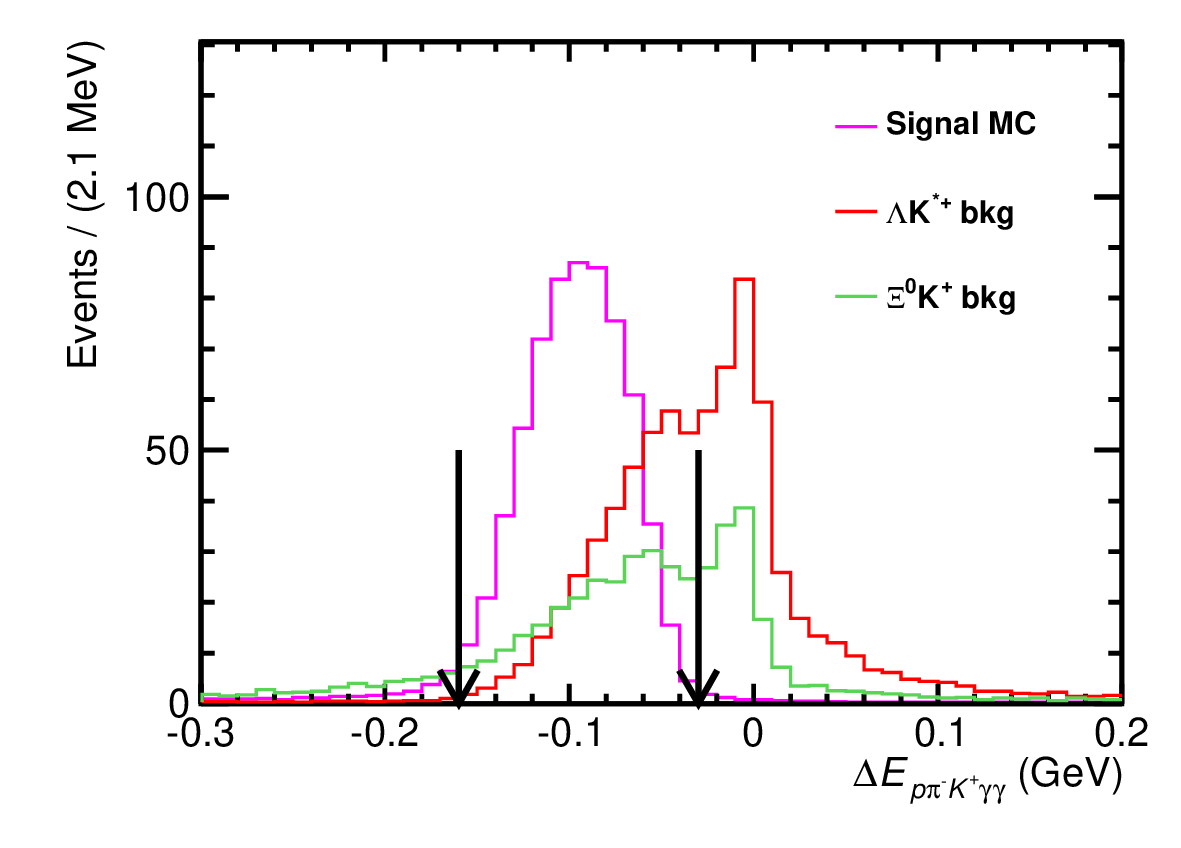}
    \includegraphics[width=0.5\textwidth]{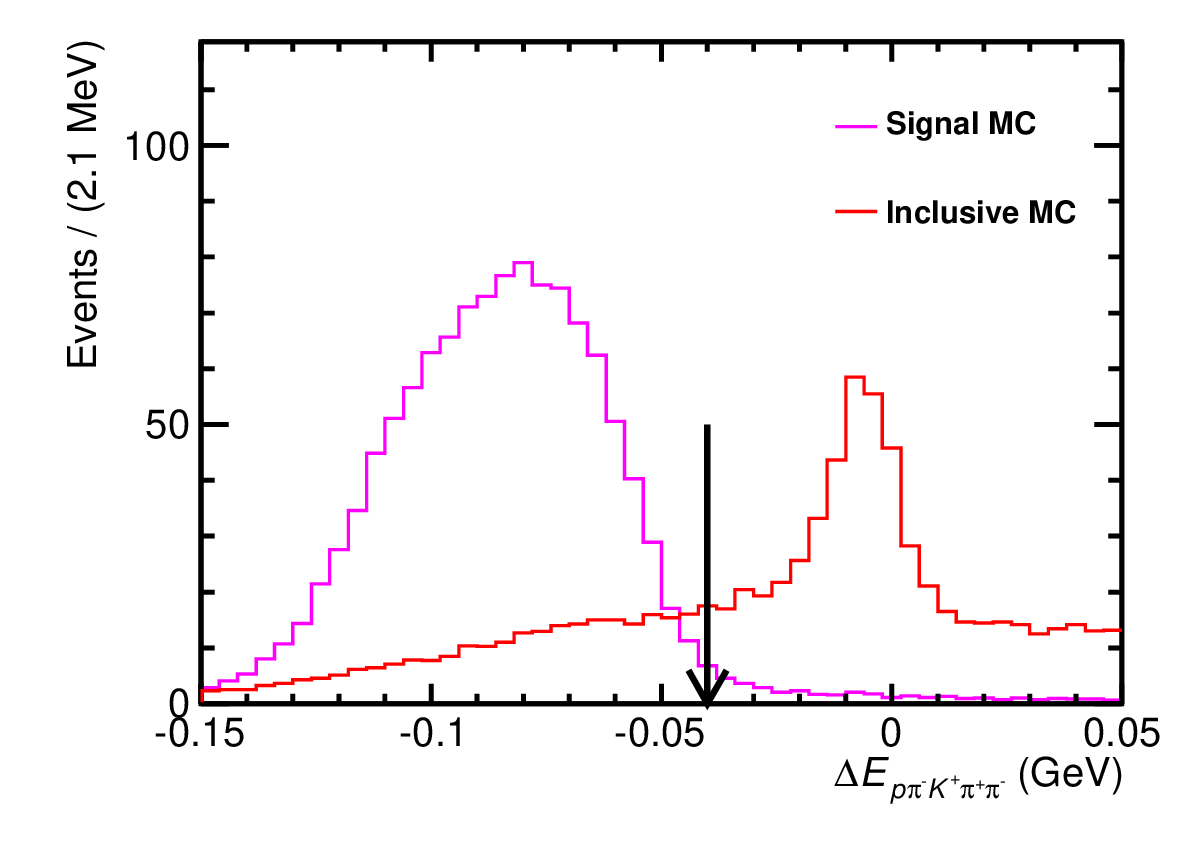}
	\figcaption{\label{Figure0} The distributions of $\Delta E_{p\pim K^{+}\gamma\gamma} (\Delta E_{p\pim K^{+}\pip\pim})$ for $\Lambda_c^+ \to \Sigma^{0}K^+\pi^{0}$($\Lambda_c^+ \to \Sigma^{0}K^+\pi^{+}\pi^{-}$). The histograms of the signal MC are normalized to make the distribution more intuitive when compared to the inclusive MC.}
\end{figure*}

 To reduce the effect from the noise produced by $\bar{p}$ in the EMC, the opening angle between photon and antiproton is required to be greater than $20^\circ$, which is obtained by optimizing the figure-of-merit defined as FOM $ = \frac{\varepsilon}{2.5+\sqrt{B}}$~\cite{2011FOM}. 
 Here, $\varepsilon$ is the signal efficiency and $B$ denotes the background yield from the inclusive MC samples. By utilizing the generic event-type analysis tool TopoAna~\cite{Topo}, the study of inclusive MC samples shows that the peaking backgrounds for $\Lambda_c^+ \to \Sigma^{0}K^+\pi^{0}$ are from the $\Lambda_c^+ \to \Xi^0 K^+$ and $\Lambda_c^+ \to \Lambda K^{*+}$ decays. These backgrounds involve one less photon in the final state than the signal process. 
 To suppress these backgrounds, we define the energy difference
 $\Delta E_{p\pim K^{+}\gamma\gamma} \equiv  E_{p} +  E_{\pim} +
 E_{K^+} + E_{\gamma1} + E_{\gamma2} - E_{\mathrm{beam}}$, where
 $E_{p}$, $E_{\pim}$, $E_{K^+}$, and $E_{\gamma1/2}$ are the energies
 of the proton, pion, kaon, and two photons, respectively while $E_{\mathrm{beam}}$ representing the beam energy. Candidate events for $\Lambda^+_c\to \Sigma^0K^+\pi^0$ are required to satisfy $-160 < \Delta E_{p\pim K^{+}\gamma\gamma} <-30$ MeV; while candidate events for $\Lambda^+_c\to \Sigma^0K^+\pi^+\pi^-$ are required to satisfy $\Delta E_{p\pim K^{+}\pip\pim} <-40$ MeV. The distributions of $\Delta E_{p\pim K^{+}\gamma\gamma} (\Delta E_{p\pim K^{+}\pip\pim})$ are shown in Fig.~\ref{Figure0}.
 
After applying the above requirements, The $\Sigma^{0}$, $K^+$ and
$\pi^0(\pi^{\pm}$) candidates are combined to reconstruct the
$\Lambda_c^{+}$. Kinematic variables, including energy difference
$\Delta E$, defined as $\Delta E \equiv E_{\mathrm{rec}-\Lambda_c^{+}}
- E_{\mathrm{beam}}$, and the beam-constrained mass $M_{\mathrm{BC}}$,
defined as $M_{\mathrm{BC}} \equiv \sqrt{E^2_{\mathrm{beam}}/^{}c^4 -
  {|\vec{p}|}^2/^{}c^2}$, are utilized to identify $\Lambda_c^{+}$
candidates. Here, $E_{\mathrm{rec}-\Lambda_c^{+}}$ and $\vec{p}$ are
the energy and momentum of $\Lambda_c^{+}$ candidate, respectively. If
there are multiple combinations satisfying these requirements in an
event, the one with the minimum $|\Delta E|$ is retained. Candidate
events for $\Lambda_{c}^{+}\to\Sigma^{0} K^{+}\pi^{0}$ and
$\Lambda^+_c\to \Sigma^0K^+\pi^+\pi^-$ are required to satisfy $\Delta
E \in [-27,~6]$ MeV and $\Delta E \in [-21,~7]$ MeV respectively, with
the ranges optimized according to the FOM. The signal efficiency and
background yield are obtained within the $M_{\rm BC}$ signal region of
$M_{\mathrm{BC}} \in [2.282,~2.291]~$GeV/$c^2$. 
 To obtain a pure signal, we have employed the truth-match method~\cite{matched}. This method involves comparing two photons in the $\pi^{0}$, one photon in the $\Sigma^{0}$, and the charged tracks $K^{\pm}$ and $\pi^{\pm}$ with their corresponding truth information. The angle $\theta_{\mathrm{truth}}$ is defined as the opening angle between each reconstructed tracks (showers) and its corresponding simulated tracks (showers). The signal shape is derived from events where $\theta_{\mathrm{truth}}$ is less than $20^\circ$ for all tracks (showers).

\begin{figure*}
    \includegraphics[width=0.5\textwidth]{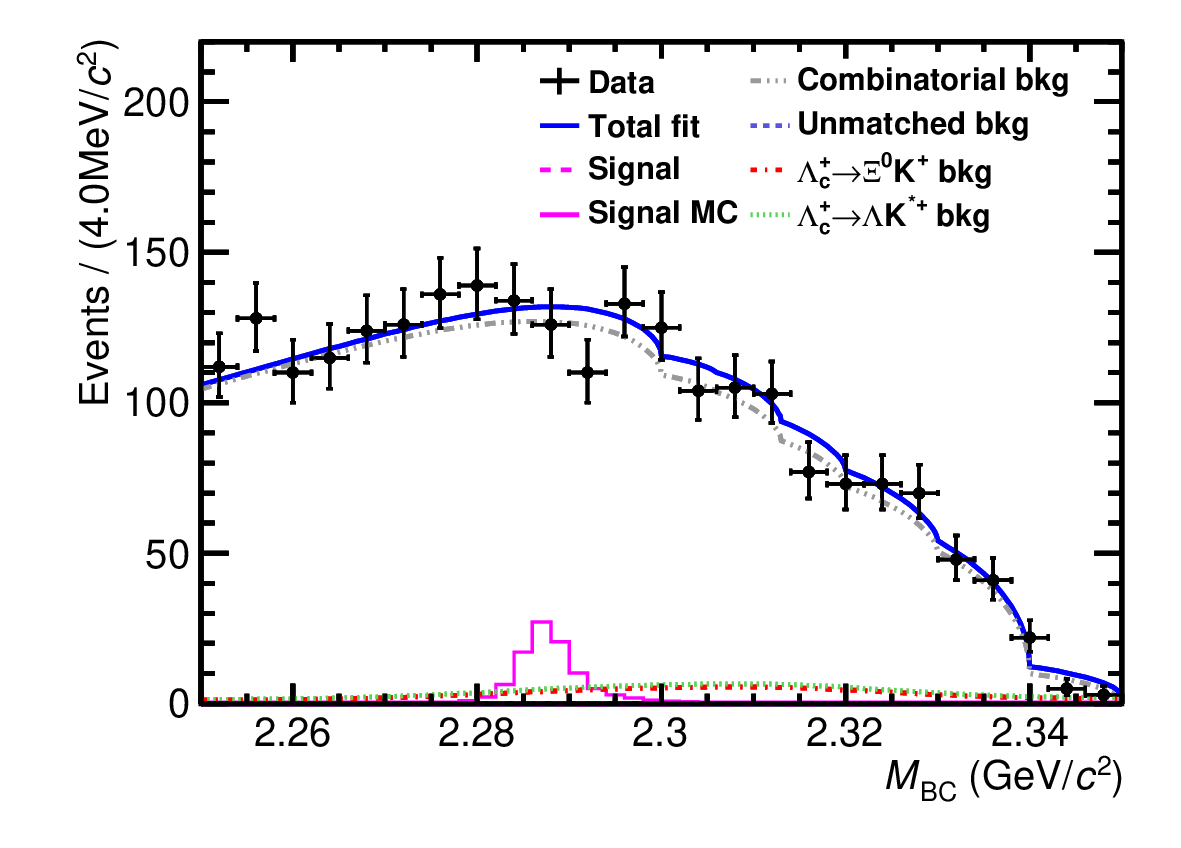}
    \includegraphics[width=0.5\textwidth]{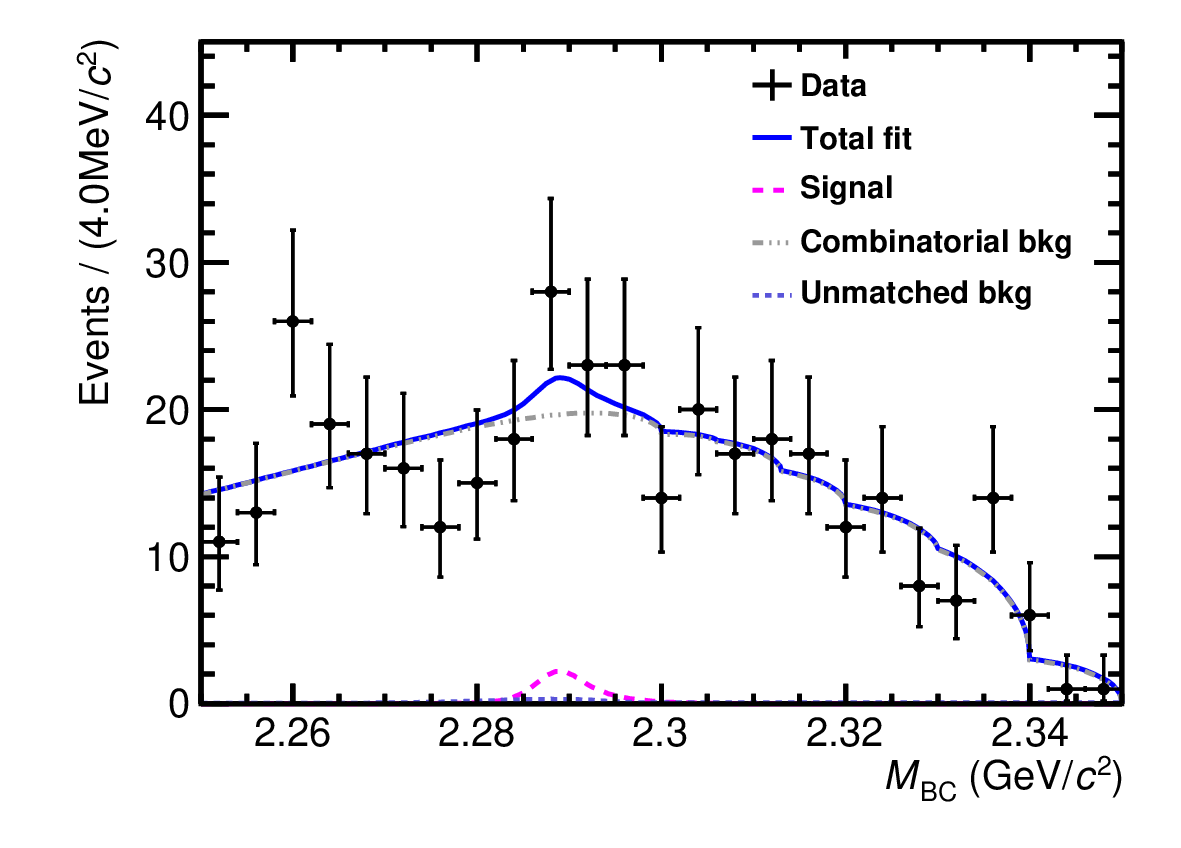}
	\figcaption{\label{Figure2} The  fit to the $M_{\rm BC}$ distributions of $\Lambda_c^+ \to \Sigma^{0}K^+\pi^{0}$(left) and $\Lambda_c^+ \to \Sigma^{0}K^+\pip\pim$(right) of the combined data. For $\Lambda_c^+ \to \Sigma^{0}K^+\pi^{0}$, the violet histograms are the signal MC samples normalized with a product BF of 1.2 $\times 10^{-3}$~\cite{Geng:2018upx}, the argus function includes seven sub-argus functions. The black point with the error bar is data, the blue solid line represents the total fit function, the gray dashed line shows the combinatorial background, the violet dash line is the signal function, the navy blue dashed line is the un-matched component, cyan dashed line is the background shape extract from $\Lambda_c^{+}\to\Lambda K^{*+}$ MC samples, and red dashed line is the background shape extract from $\Lambda_c^{+}\to\Xi^{0} K^{+}$ MC samples.}
    \label{fig:scatter}
\end{figure*}

 Table~\ref{tab:efficiency} lists the signal efficiencies obtained at different energy points.
Figure~\ref{Figure2} shows the $M_{\mathrm{BC}}$
  distributions of the simultaneous fit performed between different energy points for each of the signal decays, where no clear $\Lambda_c^+$ signals are observed. A likelihood scan method is employed after incorporating the systematic uncertainties, as discussed in the next section, to estimate the upper limits.

The absolute BF of the signal decay is determined by
\begin{equation}
\mathcal{B}^{\rm{sig}} \equiv  \frac{N^{\rm{sig}}}{2 \cdot N_{\Lambda_c^{+}\bar{\Lambda_c^{-}}}\cdot \mathcal{B}^{\mathrm{inter}} \cdot \varepsilon^{\mathrm{sig}}}, \label{eq:RBF}
\end{equation}
where $N_{\Lambda_c^{+}\bar{\Lambda}_c^{-}}$ is the total number of $\Lambda_c^{+}\bar{\Lambda}_c^{-}$ pairs,  $\varepsilon^{\mathrm{sig}}$ is the single-tag efficiency, and $\mathcal{B}^{\mathrm{inter}}$ is the product BFs of the intermediate states $\Sigma^0$, $\Lambda$ and $\pi^0$.
\begin{center}
\tabcaption{Single-tag efficiencies (\%) for $\Lambda_c^+ \to \Sigma^{0}K^+\pi^{0}$ and $\Lambda_c^+ \to \Sigma^{0}K^+\pip\pim$ at different energy points, where the uncertainties are statistical only.}
\label{tab:efficiency}
\footnotesize
\begin{tabular}{lcc}\hline
\hline
$\sqrt{s}~(\mev)$  &$\varepsilon_{\Lambda_c^+ \to \Sigma^{0}K^+\pi^{0}}$&$\varepsilon_{\Lambda_c^+ \to \Sigma^{0}K^+\pip\pim}$\\ 
\hline
4599.53 &5.17$\pm$0.04 &3.44$\pm$0.03  \\
4611.86 &4.89$\pm$0.03 &3.10$\pm$0.03   \\    
4628.00 &4.76$\pm$0.03 &3.14$\pm$0.03   \\
4640.91 &4.76$\pm$0.03 &3.21$\pm$0.03 \\
4661.24 &4.71$\pm$0.03 &3.32$\pm$0.03   \\ 
4681.92 &4.68$\pm$0.03 &3.43$\pm$0.03  \\
4698.82 &4.65$\pm$0.03 &3.44$\pm$0.03   \\
\hline
\hline
\end{tabular}
\end{center}

%Signal shapes and % Background shapes
  Since there are different distributions of background and signal events at each energy point, a simultaneous fit is performed on individual $M_{\rm BC}$ distributions. 
The BF of each signal decay is constrained to be the same value through a maximum likelihood simultaneous fit to individual $M_{\rm BC}$ distributions across seven energy points. In the fit, the signal shapes are derived from MC simulations convolved with Gaussian functions to account for the potential difference between data the MC simulations, due to imperfect modeling in MC simulation and the beam-energy spread. The control samples of  $\Lambda_c^{+}\to\Sigma^0\pi^{+}\pi^{0}$ and $\Lambda_c^{+}\to\Sigma^0\pi^{+}\pi^{+}\pi^{-}$ are used to evaluate the resolution, which have similar topologies as our signal decays.
The combinatorial backgrounds are well described by the ARGUS
function~\cite{ARGUS:1990hfq} with c.m. energy dependent endpoint
fixed at $E_{\text{beam}}$.  The remaining peaking backgrounds,
$\Lambda_c^{+}\to\Xi^{0} K^{+}$ and $\Lambda_c^{+}\to\Lambda K^{*+}$
for $\Lambda_c^+ \to \Sigma^{0}K^+\pi^{0}$, are described using
exclusive MC simulations with yields determined by the known BFs and the simulated misidentification rates as listed in Table~\ref{tab:efficiencypeak}. For $\Lambda_c^{+}\to\Sigma^0K^{+}\pi^{+}\pi^{-}$, there is no significant peaking background. 
Unmatched events, studied through the signal MC samples, exhibit a non-flat distribution. In the simultaneous fit, the yields associated with the unmatched events are determined by evaluating the ratio between the matched signal yields and the unmatched background yields, with the ratio obtained from MC simulation.\\

 \begin{center}
\tabcaption{The contamination rates (\%) after including the BFs of the secondary decays at each energy point, where the uncertainties are statistical only.}
\label{tab:efficiencypeak}
\footnotesize
\begin{tabular}{lcc}
\hline
\hline
$\sqrt{s}~(\mev) $&$\varepsilon_{\Lambda_c^+\to\Xi^0K^+}$&  $\varepsilon_{\Lambda_c^+\to\Lambda K^{*+}}$ \\
\hline
4599.53 &2.34$\pm$0.03  &2.63$\pm$0.02  \\
4611.86 &1.97$\pm$0.03  &2.59$\pm$0.02 \\
4628.00 &2.10$\pm$0.03  &2.58$\pm$0.02 \\
4640.91 &2.06$\pm$0.03  &2.58$\pm$0.02 \\
4661.24 &2.10$\pm$0.03  &2.57$\pm$0.02 \\
4681.92 &2.14$\pm$0.03  &2.56$\pm$0.02 \\
4698.82 &2.25$\pm$0.03  &2.46$\pm$0.02 \\
\hline
\hline
\end{tabular}
\end{center}

\section{Systematic uncertainty}

The systematic uncertainties in the determinations of the upper limits on the BFs are classified into two categories: additive terms and multiplicative terms.

The additive terms include the uncertainties introduced by the chosen
signal and background shapes. The uncertainty associated with the
signal shape is estimated by changing the parameters of the convolved
Gaussian functions within their uncertainties. The largest deviation
of the individual changes is taken as the uncertainty. The background
shape of the non-peaking components is changed from the ARGUS function
to be the shape extracted from the inclusive MC samples. The
uncertainty due to the fixed contribution of the peaking background
yields in the fit is investigated by varying the fixed yields within
$\pm 1 \sigma$ of the PDG BFs of individual background sources. Among
all the above terms, the case yielding the largest upper limit is
chosen for further analysis. The additive uncertainty for $\Lambda_c^+
\to \Sigma^{0}K^+\pi^+\pi^-$ is dominated by the signal shape
uncertainty, while the $\Lambda_c^+ \to
\Sigma^{0}K^+\pi^{0}$ is mainly influenced by the background shape uncertainty.

The sources of multiplicative systematic uncertainties include tracking and PID of charged particles, $\pi^0$ reconstruction, $\Lambda$ reconstruction, photon reconstruction, $\Delta E$ requirement, $\mathcal{B}^{\mathrm{inter}}$(Quoted BF), MC model, truth matching, MC statistics, $N_{\Lambda_c^+\bar\Lambda_c^-}$, $\Delta E_{\Lambda}$($\Delta E_{ p\pim K^{+}\gamma\gamma}$ and $\Delta E_{p\pim K^{+}\pip\pim}$) and $\theta_{\bar{p}\gamma}$ requirement. The total multiplicative systematic uncertainties are summarized in Table~\ref{tab:systemUncertain} and discussed in details below.

\begin{center}
\tabcaption{Multiplicative systematic uncertainties in unit of \% for the BF measurement.}
\label{tab:systemUncertain}
\footnotesize
\begin{tabular}{lcc}\hline
\hline
 Source&$\Lambda_c^+ \to \Sigma^{0}K^+\pi^{0}$&$\Lambda_c^+ \to \Sigma^{0}K^+\pip\pim$\\
 \hline
 Tracking & 1.0 & 3.0               \\
 PID & 1.0 & 3.0               \\
$\pi^0$ reconstruction &3.1  &- \\
$\Lambda$ reconstruction &2.5 &2.5 \\
Photon detection&0.5&0.5\\
$\Delta E$ requirement & 2.0 &3.7 \\
MC model & 5.5& 18.5            \\ 
$\mathcal{B}^{\mathrm{inter}}$& 0.8& 0.8\\
Truth matching & 5.5& 4.9\\
$N_{\Lambda_c^+\bar\Lambda_c^-}$&0.9&0.9\\
MC statistics & 0.5&0.3\\
$\Delta E_{\Lambda}$ requirement& 0.4&0.3\\
$\theta_{\bar{p}\gamma}$ requirement&0.1 &-\\
\hline
Total &9.2 &20.1\\
\hline
\hline
\end{tabular}
\end{center}

\begin{itemize}

\item[\bf (a)] {Tracking and PID:} The uncertainties of either PID or tracking of the charged tracks are quoted as 1.0$\%$ per track based on studies of the control sample of $e^+e^-\to K^+K^-\pi^+\pi^-$~\cite{2019Trk_and_PID}.

\item[\bf (b)] {${\pi^{0}}$ reconstruction:} The $\pi^0$ reconstruction efficiency is studied with the control samples of $\psi(3686)\to J/\psi\pi^0\pi^0$ and $e^+e^-\to \omega\pi^{0}$. The associated systematic uncertainty is assigned to be 3.1$\%$ for each $\pi^{0}$.

\item[\bf (c)] {$\Lambda$ reconstruction:} The systematic uncertainty of $\Lambda$ reconstruction is assigned to be 2.5\% by referring to the study of  $\Lambda_c^+\to\Lambda\pi^+$ in Ref.~\cite{2016Lmd}, which includes the systematics associated with reconstructing the daughter particles proton and pion.

\item[\bf (d)] {Photon reconstruction:} The systematic uncertainty due to the photon reconstruction is estimated to be 0.5\% for photon by analyzing the ISR process $e^+e^-\to \gamma\mu^+\mu^-$.

\item[\bf (e)] {$\Delta E$ requirements:} Potential differences in the $\Delta E$ distributions between data and MC simulation are studied with the control samples of 
 $\Lambda_c^+ \to \Sigma^{0}\pi^+\pi^{0}$ and $\Lambda_c^+ \to \Sigma^{0}\pi^+\pi^+\pi^-$. The differences between the nominal and alternative acceptance efficiencies, 2.0\% and 3.7\%, are taken as the systematic uncertainties for $\Lambda_c^+ \to \Sigma^{0}K^+\pi^{0}$ and $\Lambda_c^+ \to \Sigma^{0}K^+\pi^+\pi^-$, respectively. 

\item[\bf (f)] {MC model:} The systematic uncertainties associated with the MC model are evaluated with alternative signal MC samples for $\Lambda_c^+ \to \Sigma^{0}K^+\pi^{0}$ and $\Lambda_c^+ \to \Sigma^{0}K^+\pi^+\pi^-$. These samples are generated as  
$\Lambda_c^+\to\Lambda(1405) K^+$, with $\Lambda(1405)\to \Sigma^{0}\pi^0$ via the PHSP model,
and $\Lambda_c^+\to\Sigma^0\pip
K^{*}$, with $K^{*}\to K^+\pim$ also simulated in the PHSP model. The differences between the efficiencies of these alternative models and the nominal model are taken as the systematic uncertainties, which are $5.5\%$ for $\Lambda_c^+ \to \Sigma^{0}K^+\pi^{0}$ and $18.5\%$ for $\Lambda_c^+ \to \Sigma^{0}K^+\pi^+\pi^-$, respectively.

\item[\bf (g)] {$\mathcal{B}^{\mathrm{inter}}$:} The BFs of
  $\Sigma^{0}\to\Lambda\gamma,~ \Lambda\to p\pim ~\mathrm{and}~
  \pi^{0}\to\gamma\gamma$, are quoted from the
  PDG~\cite{ParticleDataGroup:2024cfk}.  The uncertainties of these
  known BFs add up to a total uncertainty of 0.8\%.

\item[\bf (h)] {Truth matching:} To estimate the uncertainty caused by the angle cut in deriving the signal MC shape, we loosen the cut by $5^\circ$ for each angle. The differences between the nominal and new efficiencies are taken as the systematic uncertainties,  which are $5.5\%$ for $\Lambda_c^+ \to \Sigma^{0}K^+\pi^{0}$ and $4.9\%$ for $\Lambda_c^+ \to \Sigma^{0}K^+\pi^+\pi^-$.

\item[\bf (i)] {$N_{\Lambda_c^+\bar\Lambda_c^-}$:} The uncertainty of $N_{\Lambda_c^+\bar\Lambda_c^-}$ is quoted from Refs.~\cite{BESIII:2022Lambdac1,BESIII:2022Lambdac2}. Its effect on the BF measurement, 0.9\%, is assigned as the systematic uncertainty for both decays.

\item[\bf (j)] {MC statistics:} The uncertainties due to limited MC statistics are 0.5$\%$ for $\Lambda_c^+ \to \Sigma^{0}K^+\pi^{0}$ and 0.3$\%$ for $\Lambda_c^+ \to \Sigma^{0}K^+\pi^+\pi^-$.

\item[\bf (k)] {$\Delta E_{\Lambda}$ requirement:} The uncertainty due to the $\Delta E_{\Lambda}$ requirement is estimated with the control samples of $\Lambda_c^+ \to \Sigma^{0}\pi^+\pi^{0}$ and $\Lambda_c^+ \to \Sigma^{0}\pi^+\pi^{+}\pi^{-}$. The maximum changes of the acceptance efficiencies between data and MC simulation by varying the $\Delta E_\Lambda$ requirement by $\pm0.05$~GeV are taken as the systematic uncertainties, which are $0.4\%$ and $0.3\%$  for $\Lambda_c^+ \to \Sigma^{0}K^+\pi^{0}$ and $\Lambda_c^+ \to \Sigma^{0}K^+\pi^+\pi^-$, respectively. 

\item[\bf (l)] {$\theta_{\bar{p}\gamma}$ requirement:} The systematic uncertainty from the $\theta_{\bar{p}\gamma}$ requirement for $\Lambda^+_c\to \Sigma^0K^+\pi^0$ is  estimated with the control sample of $\Lambda^+_c\to \Sigma^0\pi^+\pi^0$. The maximum change of the acceptance efficiencies between data and MC simulation after varying the $\theta_{\bar p\gamma}$ requirement by $\pm5^\circ$, 0.1\%, is assigned as the systematic uncertainty.  
\end{itemize}

\section{Results}
  The fit result is consistent with a background-only hypothesis of $\Lambda_c^+ \to \Sigma^{0}K^+\pi^{0}$ and $\Lambda_c^+ \to \Sigma^{0}K^+\pip\pim$ , and the upper limits on their BFs are determined.
The distributions of raw likelihoods versus individual BFs are shown
as the blue dashed curves in Fig.~\ref{fig:prob}. Each curve is then
convolved with a Gaussian function with zero mean and width set as the corresponding multiplicative systematic uncertainty, according to Refs.~\cite{K.Stenson:2006,cpc:up}. The updated likelihood distributions are shown as the red solid lines in Fig.~\ref{fig:prob}. By integrating the red solid curves from zero to 90\% of the physical region,
the upper limits on the BFs at the 90\% C.L. are set to be

\begin{equation*}
\begin{aligned}
&{\mathcal B}(\Lambda_c^+ \to \Sigma^{0}K^+\pi^{0})< 5.0\times 10^{-4},\\
&{\mathcal B}(\Lambda_c^+ \to \Sigma^{0}K^+\pi^{+}\pi^-)< 6.5\times 10^{-4}.
\end{aligned}
\end{equation*}

\section{Summary}
Based on 4.5 fb$^{-1}$ of $\ee$ annihilation data collected at
c.m. energies between 4599.53 MeV and 4698.82 MeV with the BESIII
detector at the BEPCII collider, we present the studies of the singly
Cabibbo-suppressed hadronic decays $\Lambda_c^+ \to
\Sigma^{0}K^+\pi^{0}$ and
$\Lambda_c^{+}\to\Sigma^0K^{+}\pi^{+}\pi^{-}$ using a single-tag
method. The upper limits on their BFs at the $90\%$ C.L. are set to be
$5.0\times 10^{-4}$ for $\Lambda_c^+ \to \Sigma^{0}K^+\pi^{0}$ and
$6.5\times 10^{-4}$ for
$\Lambda_c^{+}\to\Sigma^0K^{+}\pi^{+}\pi^{-}$. The upper limit of the
BF of $\Lambda^+_c\to \Sigma^0K^+\pi^0$ is more stringent than the
previous BESIII measurement using a double-tag method~\cite{BESIII:2023SKP}.
The upper limit of the BF of $\Lambda_c^+ \to \Sigma^{0}K^+\pi^{0}$ is compatible with the theoretical predictions\cite{Cen:2019ims,2024LMD} and could not rule out any of them, as shown in Table~ \ref{tab:total}. For $\Lambda_c^{+}\to\Sigma^0K^{+}\pi^{+}\pi^{-}$, the upper limit is less stringent than the BaBar result~\cite{BaBar:2006eah}.
These results provide valuable information for understanding the dynamics of charmed baryon decays and improving theoretical models. The sensitivities to these two decays could be further improved with a larger data sample at BESIII~\cite{2020BESIII} in the near future. 

 \begin{center}
\centering
\includegraphics[width=0.45\textwidth]{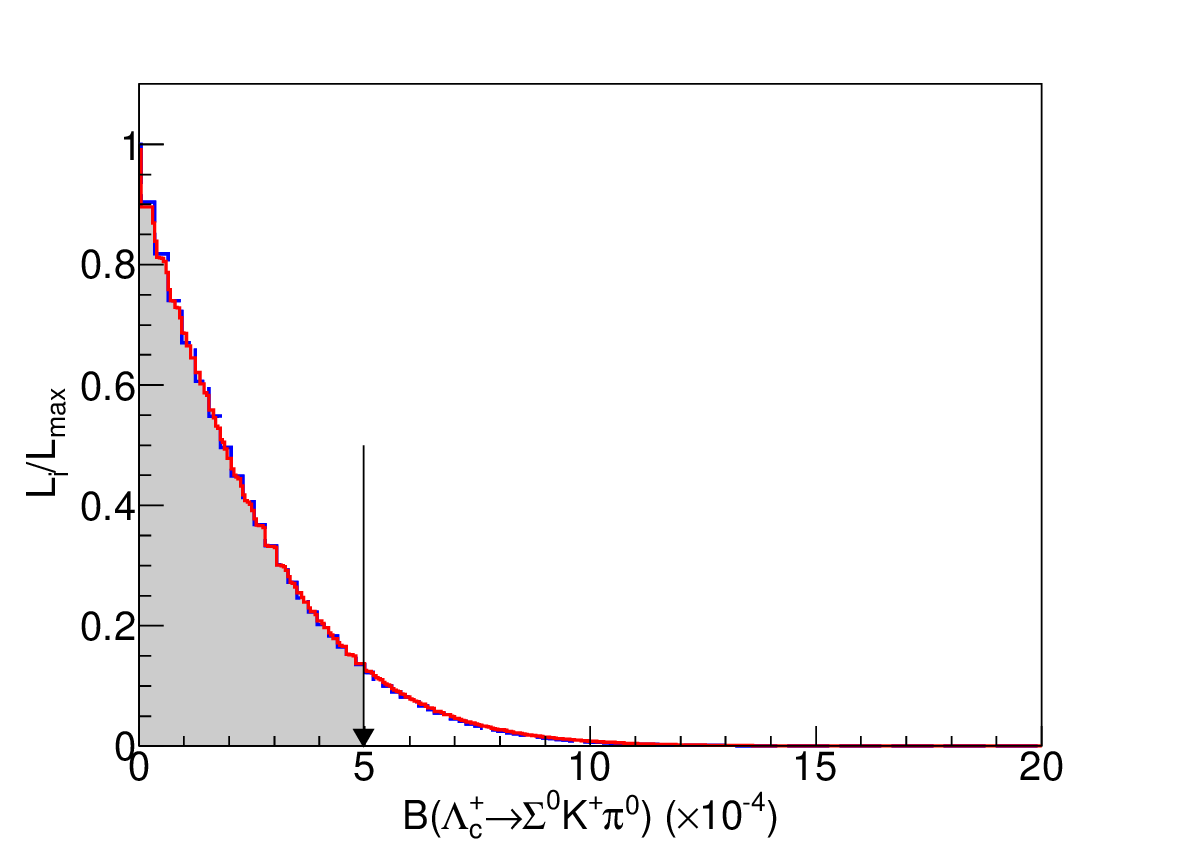}
 \includegraphics[width=0.45\textwidth]{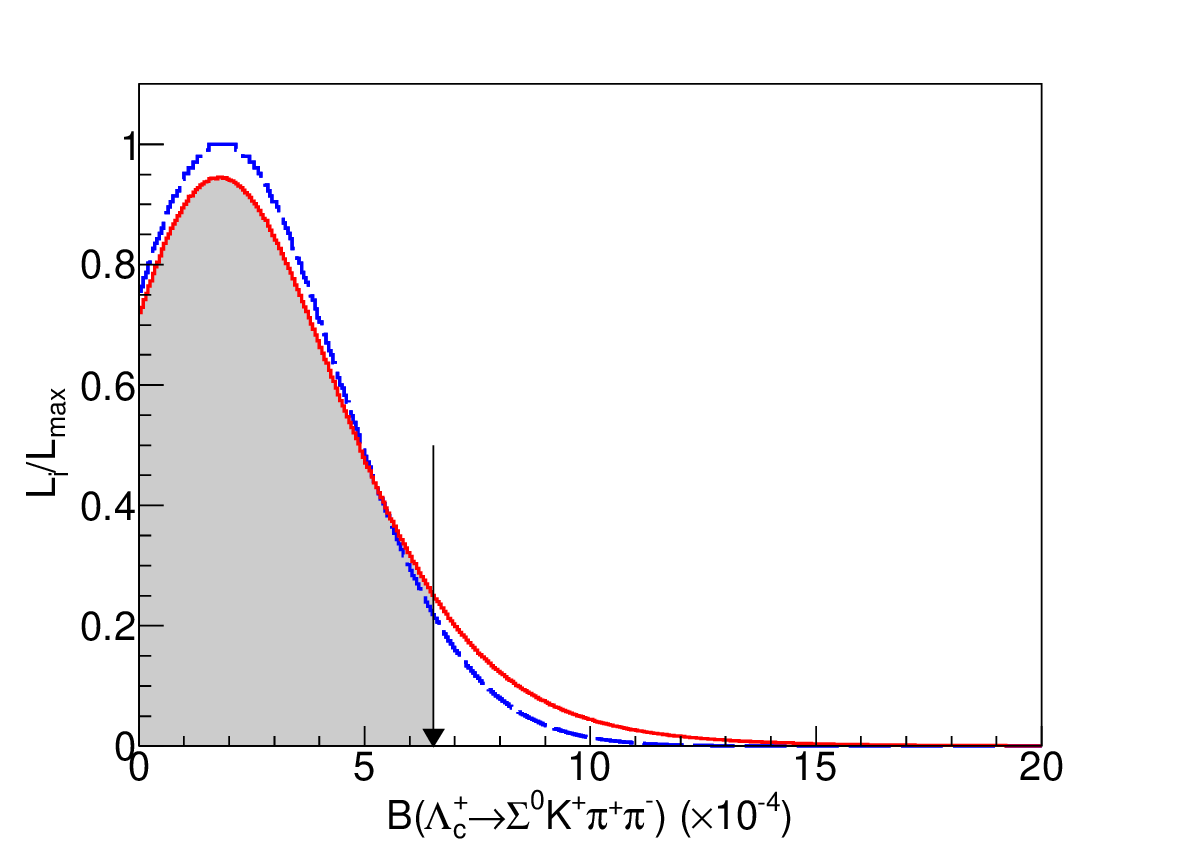}
\figcaption{\label{Figure4} The distributions of the likelihoods versus the BFs of $\Lambda_c^+ \to \Sigma^{0}K^+\pi^{0}$(top) and $\Lambda_c^+ \to \Sigma^{0}K^+\pip\pim$(bottom). The results obtained with and without incorporating the systematic uncertainties are shown in the red solid and blue dashed curves, respectively. The black arrows shows the results corresponding to the 90\% C.L..}
\label{fig:prob}
\end{center}

\begin{table*}
\tabcaption{\label{tab:total} Comparison of the experimental measurements of $\Lambda_c^+ \to \Sigma^{0}K^+\pi^{0}$ and  $\Lambda_c^{+}\to\Sigma^0K^{+}\pi^{+}\pi^{-}$ obtained in this work, and those of BaBar and BESIII (single-tag) as well as theoretical predictions.}
\centering
\footnotesize
\begin{tabular}{ccc} 
\hline
\hline
Decay mode &$\Lambda_{c}^{+}\to\Sigma^{0} K^{+}\pi^{0}$&$\Lambda_c^{+}\to\Sigma^0K^{+}\pi^{+}\pi^{-}$    \\
\hline
C.Q.Geng {\it et al.}~\cite{Geng:2018upx}& $(1.2\pm0.3) \times 10^{-3}$&- \\
\hline
J.Y.Cen {\it et al.}~\cite{Cen:2019ims} &$(0.8\pm0.2) \times 10^{-3}$&-\\
\hline
C.Q.Geng {\it et al.}~\cite{2024LMD}& $(8.2\pm1.4) \times 10^{-4}$&-\\
\hline
M.Gronau {\it et al.}~\cite{2018MGronau} &$(2.1\pm0.6) \times 10^{-3}$&- \\
\hline
BESIII~(double-tag)~\cite{BESIII:2023SKP}&$\textless 1.8 \times 10^{-3}$&-\\
\hline
BaBar experiment~\cite{BaBar:2006eah}&-&$ \textless 2.5 \times 10^{-4}$\\
\hline
BESIII~(single-tag)~&$\textless 5.0 \times 10^{-4}$&$\textless 6.5 \times 10^{-4}$\\
\hline
\hline
\end{tabular}
\end{table*}

\acknowledgments{
The BESIII collaboration thanks the staff of BEPCII and the IHEP computing center for their strong support.
}

\end{multicols}

\vspace{-1mm}
\centerline{\rule{80mm}{0.1pt}}
\vspace{2mm}

\begin{multicols}{2}

\end{multicols}

\clearpage
\end{CJK*}
\end{document}

%% file: authorlist_2024-10-09.tex
%\author{Author list}
\begin{small}
\begin{center}
M.~Ablikim(麦迪娜)$^{1}$, M.~N.~Achasov$^{4,c}$, P.~Adlarson$^{76}$, X.~C.~Ai(艾小聪)$^{81}$, R.~Aliberti$^{35}$, A.~Amoroso$^{75A,75C}$, Q.~An(安琪)$^{72,58,a}$, Y.~Bai(白羽)$^{57}$, O.~Bakina$^{36}$, Y.~Ban(班勇)$^{46,h}$
% ...
\\
\vspace{0.2cm}
(BESIII Collaboration)\\
\vspace{0.2cm} {\it
$^{1}$ Institute of High Energy Physics, Beijing 100049, People's Republic of China\\
% ...
$^{4}$ Budker Institute of Nuclear Physics SB RAS (BINP), Novosibirsk 630090, Russia\\
% ...
$^{35}$ Johannes Gutenberg University of Mainz, Johann-Joachim-Becher-Weg 45, D-55099 Mainz, Germany\\
$^{36}$ Joint Institute for Nuclear Research, 141980 Dubna, Moscow region, Russia\\
% ...
$^{46}$ Peking University, Beijing 100871, People's Republic of China\\
% ...
$^{57}$ Southeast University, Nanjing 211100, People's Republic of China\\
$^{58}$ State Key Laboratory of Particle Detection and Electronics, Beijing 100049, Hefei 230026, People's Republic of China\\
% ...
$^{72}$ University of Science and Technology of China, Hefei 230026, People's Republic of China\\
% ...
$^{75}$ University of Turin and INFN, (A)University of Turin, I-10125, Turin, Italy; (B)University of Eastern Piedmont, I-15121, Alessandria, Italy; (C)INFN, I-10125, Turin, Italy\\
$^{76}$ Uppsala University, Box 516, SE-75120 Uppsala, Sweden\\
% ...
$^{81}$ Zhengzhou University, Zhengzhou 450001, People's Republic of China\\
\vspace{0.2cm}
$^{a}$ Deceased\\
% ...
$^{c}$ Also at the Novosibirsk State University, Novosibirsk, 630090, Russia\\
% ...
$^{h}$ Also at State Key Laboratory of Nuclear Physics and Technology, Peking University, Beijing 100871, People's Republic of China\\
% ...
}\end{center}

\vspace{0.4cm}
\end{small}